\documentstyle[12pt,procl]{article}

\input epsf

\newcommand{\figGluAB}
{
\begin{figure}[hbt]
\epsfxsize=\hsize
\centerline{\epsfbox{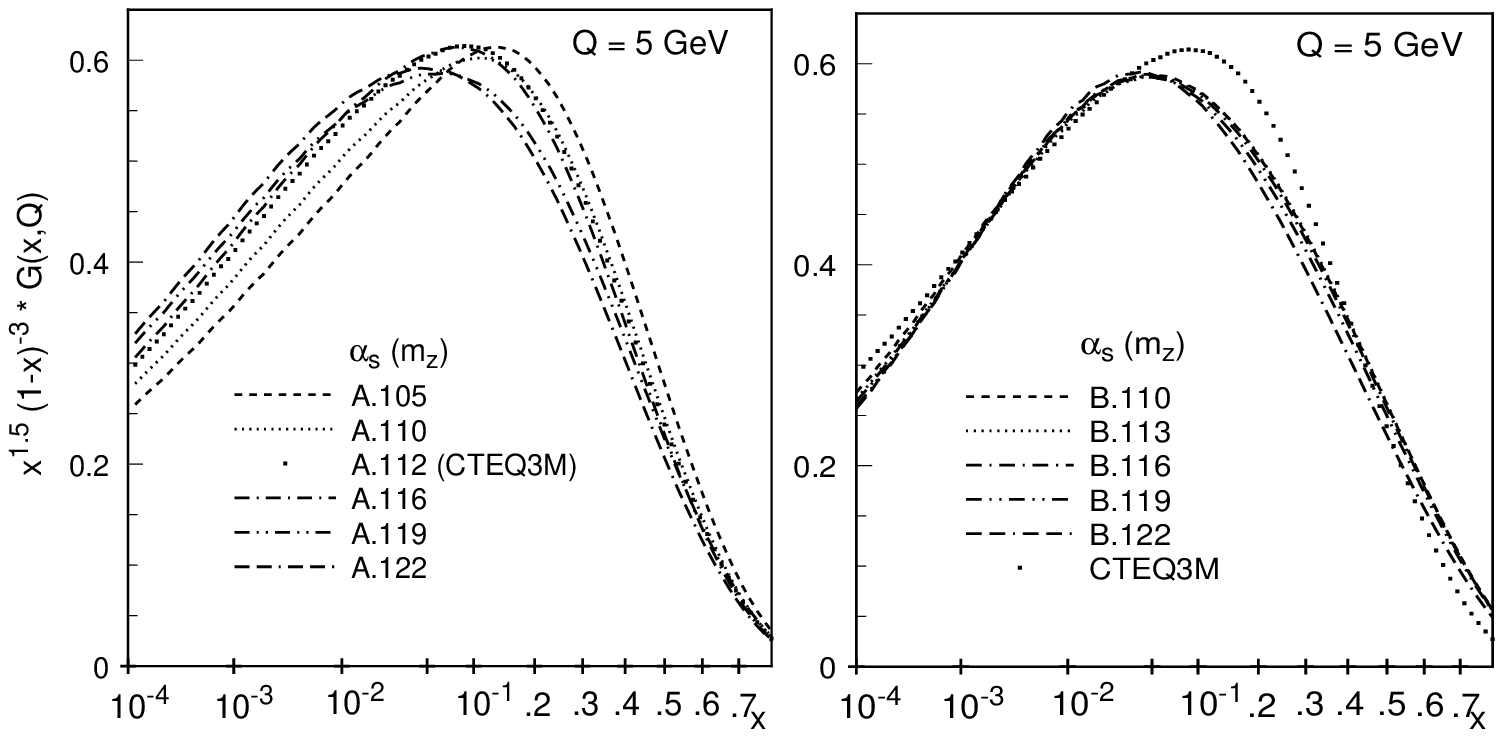}}
  \caption{Comparison of gluons obtained with pre-1995 DIS
data (A-series) with those using current DIS data (B-series). The ``minimal''
parametrization of $G(x,Q)$ is used.}
  \label{figGluAB}
\end{figure}
}

\newcommand{\figGluCdA}
{
\begin{figure}[hbt]
\epsfxsize=\hsize
\centerline{\epsfbox{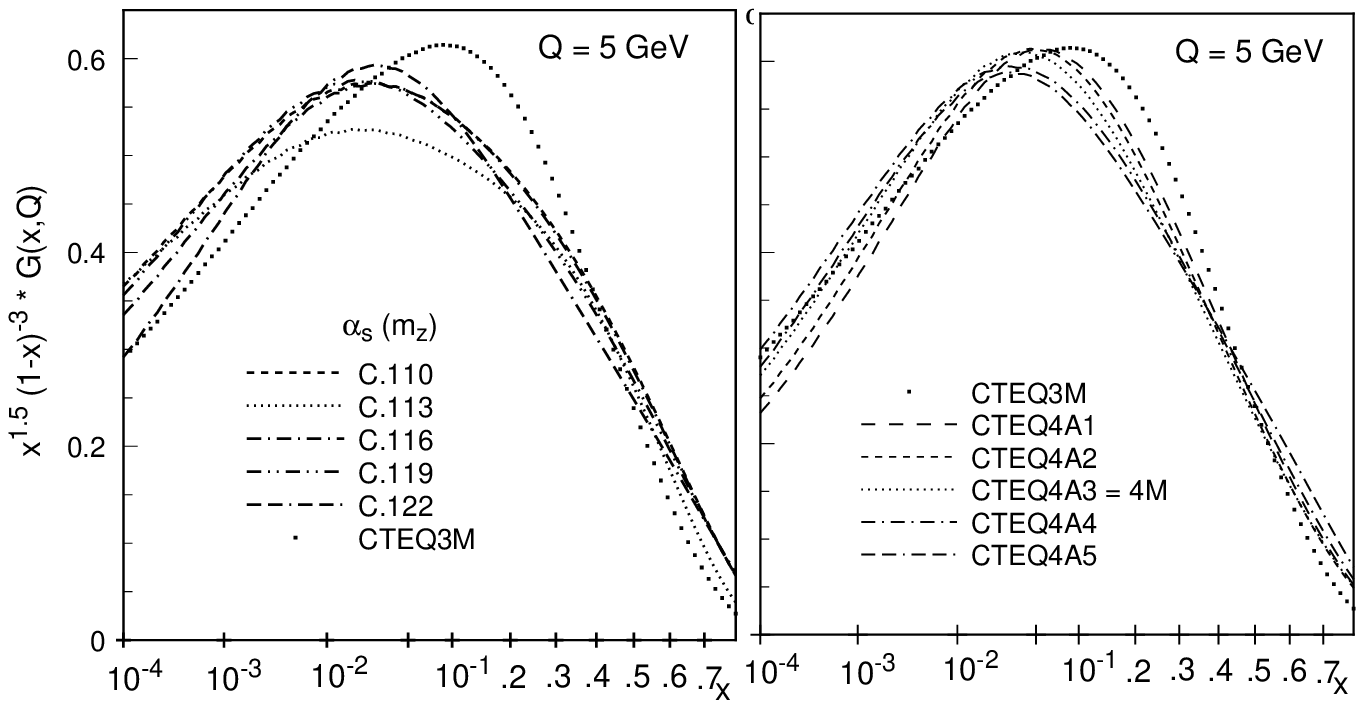}}
  \caption{Comparison of gluons obtained without jet
data (C-series) with those obtained with jet data (CTEQ4A series). The
``minimal+2'' parametrization of $G(x,Q)$ is used.}
  \label{figGluC4A}
\end{figure}
}

\newcommand{\figJetFit}
{
\begin{figure}[hbt]
\epsfxsize=\hsize
\centerline{\epsfbox{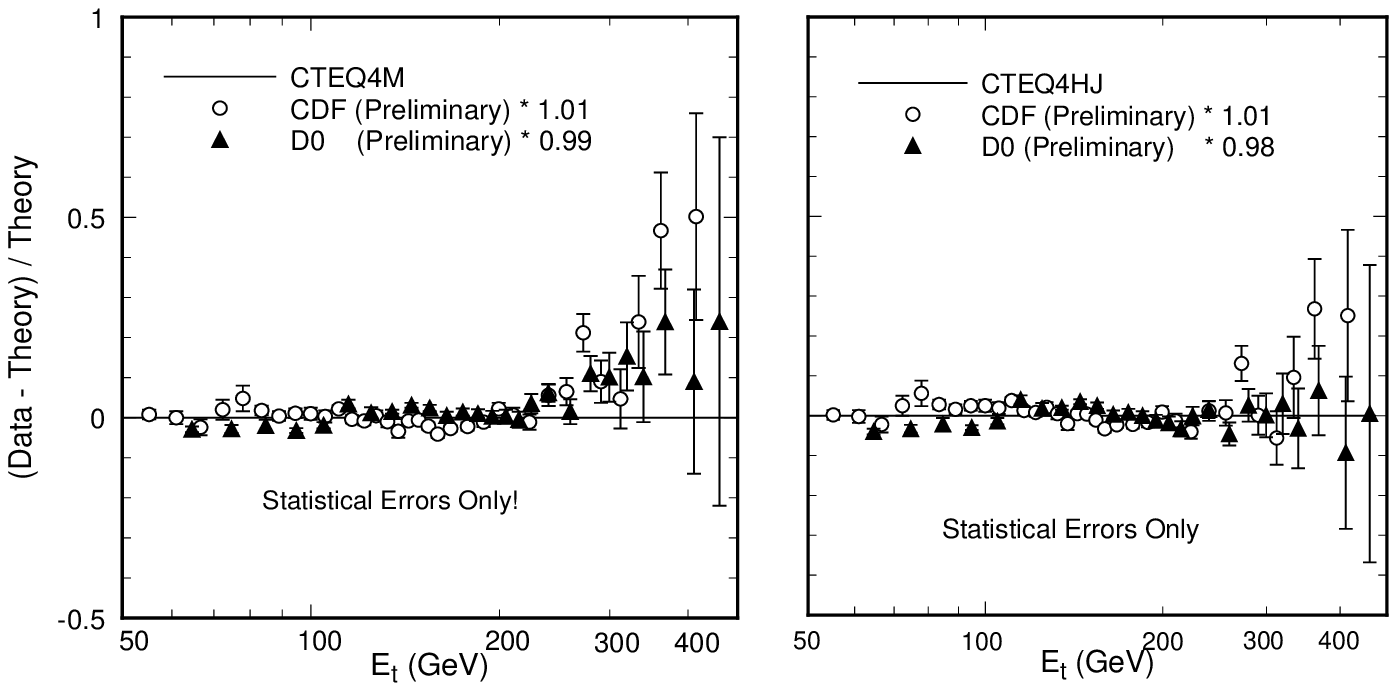}}
  \caption{CDF and D0 data compared to NLO QCD using CTEQ4M and CTEQ4HJ.}
  \label{figJetFit}
\end{figure}
}

\newcommand{\figChisqAlfa}
{
\begin{figure}[hbt]
\epsfxsize=\hsize
\centerline{\epsfbox{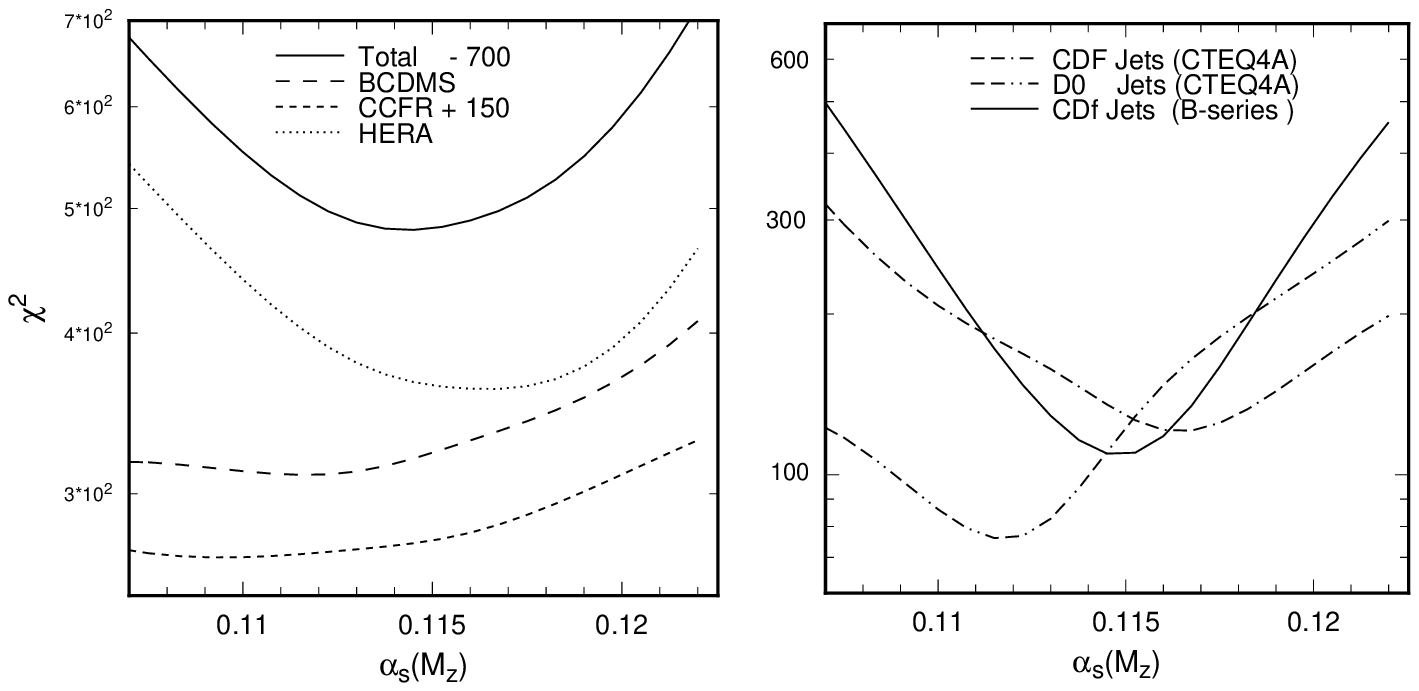}}
  \caption{$\chi^2$ vs. $\alpha_s(M_Z)$ for global fits based on current
experiments: (a) BCDMS, CCFR, combined HERA collider experiments, and total of
DIS+DY; (b) $\chi^2_{stat.err.only}$ of CDF and D0 jets, using CTEQ4A and
B-series parton distributions.}
  \label{figChisqAlfa}
\end{figure}
}

\newcommand{\tblFits}
{
\begin{table}[hbt]

\caption{(a) Several series of global fits on which the physics discussions are
based. ``New DIS'' refers to DIS data becoming available since 1995.
Minimal ``m'' and minimal+2 ``m+2'' parametrizations are explained in the text;
(b) New CTEQ4 parton distribution sets}

\label{tblFits}
\begin{center}
\begin{tabular}{|c|c|c|c|}
\hline
Series & New  & Incl. & param \\
& \smcap{DIS}  & Jets &   \\ \hline\hline
A &  &  & m  \\ \hline
B & x &  & m  \\ \hline
C & x &  & m+2  \\ \hline
\smcap{CTEQ4A} & x & x & m+2 \\ \hline
Q$_{cut}$ & x & x & m \\ \hline
\end{tabular}
\hfill
\begin{tabular}{|c|c|c|}
\hline
\smcap{PDF} set & feature & $\alpha _s(m_z)$ \\ \hline\hline
\smcap{CTEQ4M} & $\overline{MS}$ sch. & 0.116 \\ \hline
\smcap{CTEQ4D} & \smcap{DIS} sch. & 0.116 \\ \hline
\smcap{CTEQ4L} & \smcap{LO} & 0.132 \\ \hline\hline
\smcap{CTEQ4A1-5} & $\alpha_s$ series & .110-.122 \\ \hline
\smcap{CTEQ4HJ} & Hi-Jet & 0.116\\ \hline
\smcap{CTEQ4LQ} & Low Q$_0$ & 0.114 \\ \hline
\end{tabular}
\end{center}

\end{table}
}

\newcommand{\tblChiSq}
{
\begin{table}[htb]
\begin{center}

\caption{Total $\chi^2$ ($\chi^2/point$) values and their distribution among
the DIS and DY experiments for current generation of parton distributions
compared to previous one. }

\label{tblChiSq}
\begin{tabular}{|c|c||c|c|c|c|c|}
\hline
 Expt.     & \#pts &  \smcap{CTEQ4M}     &  \smcap{CTEQ4HJ}     &  \smcap{MRSJ}
   &  CTEQ3M        \\
\hline\hline
\smcap{DIS-F.T.}   &   817  &  855.2(1.05) &  884.3(1.08) & 1024.1(1.25) &
937.4(1.15) \\ \hline
\smcap{DIS-HERA}   &   351  &  362.3(1.03) &  352.9(1.01) &  362.0(1.03) &
769.7(2.19) \\ \hline
\smcap{DY} rel.    &   129  &  102.6(0.80) &  105.5(0.82) &  103.6(0.80) &
96.0(0.74) \\ \hline
\hline
 Total     &  1297  &   1320      &   1343      &   1490      &   1803      \\
\hline
\end{tabular}
\end{center}
\end{table}
}

\newcommand{\smcap}[1]{{\small #1}}

\begin{document}

\title{Global QCD Analysis, the Gluon Distribution, and High $E_t$
Inclusive Jet Data\footnote
{To appear in Proceedings of Workshop on {\em Deep Inelastic Scattering 
and Related Phenomenoa}, Rome, Italy, April, 1996. This work is done in collaboration with H.L. Lai, S. Kuhlmann, J. Huston, J.
Owens, D. Soper, and H. Weerts. It is supported by DOE and NSF.}
}

\author{Wu-Ki Tung}

\address{Michigan State University, E. Lansing, MI, USA}

\maketitle\abstracts{
We report on an extensive global {\footnotesize QCD} analysis of new
{\footnotesize DIS} and hadronic inclusive jet production data emphasizing the
impact of these recent data on the determination of the gluon distribution, and
on the interpretation of the high $E_t$ jets highlighted by the {\footnotesize
CDF} collaboration. This analysis results in (i) a better handle on the range
of uncertainty of the gluon distribution, (ii) a new generation of
{\footnotesize CTEQ} parton distributions which incorporates this uncertainty,
(iii) a viable scenario for accommodating the high $E_t$ jets in the
conventional p{\footnotesize QCD} framework, and (iv) a systematic study of the
sensitivity of the various hard processes to $\alpha_s$ and the consistency of
$\alpha_s$ determination in global analysis.
}

\section{Introduction} \label{sec:Intro}

Global \smcap{QCD} analysis of lepton-hadron and hadron-hadron hard processes
has made steady progress in testing the consistency of perturbative \smcap{QCD}
(p\smcap{QCD}) with global data and in yielding increasingly detailed
information on the universal parton distributions inside hadrons.  The quark
distributions inside the nucleon have been quite well determined from precise
\smcap{DIS} and other processes in the last few years. Recent emphasis has
therefore been mainly on the more elusive gluon, $G(x,Q)$.\cite{one} The
quantitative determination of $G(x,Q)$ is closely related to the measurement of
$\alpha_s$: the two quantities are strongly coupled since the gluon mediates
the strong force.
Direct photon production has long been regarded as potentially
the most useful source of information on $G(x,Q)$. However, in recent years, it
has been realized that a number of large
theoretical uncertainties (e.g. significant scale dependence, and $k_t$
broadening of initial state partons due to gluon radiation)\cite{CtqDph,jet1}
need to be brought under control before direct photon data can place a tight
constraint on the gluon distribution.

Inclusive jet production in hadron-hadron collisions is very sensitive to
$\alpha_s$ and $G(x,Q)$. \smcap{NLO QCD} calculations of jet cross-sections
have reached a mature stage.\cite{JetTh} Many issues relating to jet definition
(which is important for comparing theory with experiment) encountered in
earlier stages of jet analysis have been extensively studied and are better
understood. For the moderate to large $E_t$ range, the scale dependence of the
\smcap{NLO} inclusive jet cross section has been found to be relatively small.
Recently, good data on single jet production have become available over a wide
range of transverse energy, $15$ GeV $<E_t<450$ GeV.\cite{JetExp}
Thus, it is natural to incorporate inclusive jet data in a global \smcap{QCD}
analysis. \smcap{DIS} data in the small-$x$ region has steadily improved since
the advent of \smcap{HERA}. The $(x,Q)$ dependence of the measured structure
functions is sensitive to the indirect influence of gluon since $G(x,Q)$ is
about an order of magnitude larger than the quark distributions at small-$x$.
Thus, we expect both the new jet data and the recent high precision data from
the 1994 run of \smcap{HERA} to play an important role in placing constraints
on $G(x,Q)$.

\section{Global Analyses}\label{sec:GA}

Our global \smcap{QCD} analysis incorporates fixed-target \smcap{DIS} data
\cite{one} of \smcap{BCDMS, CCFR, NMC, E665}; lepton pair production
(\smcap{DY}) data of \smcap{E605, CDF}; direct photon data of \smcap{WA70,
UA6}; DY asymmetry data of \smcap{NA51}; W-lepton asymmetry data of
\smcap{CDF}; in addition to the recent \smcap{DIS} data \cite{DisExp} of
\smcap{NMC, H1, ZEUS} and the hadronic inclusive jet data \cite{JetExp} of
\smcap{CDF} and \smcap{D0}. The total number of 1300 data points included (with
$Q>2$ GeV) cover a wide triangular region on a kinematic map with $log(1/x)$
and $log Q$ as axes which is anchored by \smcap{HERA} data at small $x=10^{-4}$
in one corner and Tevatron jet data at high $Q=450$ GeV in the other.
All these processes are treated consistently in \smcap{NLO} p\smcap{QCD}.

\tblFits
In addition to finding good fits to these data, a systematic effort has been
made to investigate separately the impact of the new \smcap{DIS} and jet data
and their mutual compatibility, and to assess the range of remaining
uncertainty of the gluon distribution.  To this end, several series of global
fits, listed in Table~\ref{tblFits}, have been carried out to explore the
influence on the gluon distribution due to (i) variation of the strong coupling
$\alpha_s(M_z)$ (henceforth abbreviated simply as $\alpha_s$) within the
currently accepted range of $0.116 \pm 0.006$, (ii) choice of parametrization
of the initial $G(x,Q)$ (at $Q=Q_0=1.6$ GeV) which we take to be either the
``minimal'' form \cite{cteq4} of \smcap{CTEQ3} or a more general ``minimal+2''
form \cite{cteq4} used in \smcap{CTEQ2} and recent \smcap{MRS} parton sets, and
(iii) choice of lower cutoff in $Q$ (referred to as  $Q_{cut}$) of experimental
data to be included in the analysis.

Details of these studies will be reported elsewhere.\cite{cteq4} Here we give a
very brief summary of the main results. (a) By comparing the A- and B-series of
fits, cf. Table~\ref{tblFits}, we found that recent \smcap{DIS} data
\cite{DisExp} of \smcap{NMC, E665, H1} and \smcap{ZEUS} considerably narrow
down parton distributions compared to the \smcap{CTEQ3} and \smcap{MRSA} sets,
especially if the {\em minimal} parametrization form for $G(x,Q)$ is adopted
(B-series). Cf. Fig.~\ref{figGluAB}; \figGluAB (b) By comparing the B- and
C-series of fits, we found that the new \smcap{DIS} data do not fully constrain
$G(x,Q)$ in the more generalized parametrization (C-series); (c) By comparing
the new inclusive jet data of \smcap{CDF} and \smcap{D0} with theory
predictions based on \smcap{PDF}'s determined in the B- and C-series, we found
that they agree quite well, implying an impressive consistency between the new
jet production process and the others within the p\smcap{QCD} framework; (d) By
adding the jet data to the global fit, we improve on the parton distributions
found in the B- and C-series. The jet data has a significant effect in more
fully constraining $G(x,Q)$ in the case of the general {\em (minimal+2)}
parametrization compared to the C-series, cf. Fig.~\ref{figGluC4A};
(e) By varying the $Q_{cut}$ we obtain the range of change in $G(x,Q)$ and
found it to be less than that due to the variation in $\alpha_s$.

As the result of these detailed study, we arrive at a new generation of
\smcap{CTEQ4} parton distributions which consists of (i) the three standard
sets designated as \smcap{CTEQ4M} (\mbox{\small {$\overline {\rm MS}$}}),
\smcap{CTEQ4D} ({\protect \small DIS}), and \smcap{CTEQ4L} (leading order);
(ii) a series, \smcap{CTEQ4A1-5}, that gives a range of parton distributions
with corresponding $\alpha_s$'s; and (iii) a set with a low starting value of
$Q$, \smcap{CTEQ4LQ}. The \smcap{CTEQ4A} series represents the improved
C-series mentioned in the above paragraph, cf. Table~\ref{tblFits} and
Fig.~\ref{figGluC4A}. These parton distributions sets give very good fits to
the full range of data. The central member of the \smcap{CTEQ4A} series,
\smcap{CTEQ4A3}, which gives the best overall fit is chosen to be the standard
\smcap{CTEQ4M}. The $\alpha_s$ value for this set, as well as its {\protect
\small DIS} counterpart \smcap{CTEQ4D}, is 0.116, corresponding to
$\Lambda_5=202$ MeV. For all these sets, $Q_0=1.6$ GeV, except for
\smcap{CTEQ4LQ} which has $Q_0=0.7$ GeV.
\figGluCdA

Space does not allow a presentation of the comparison between data and fits.
The quality of these fits and the progress that has been made in this new round
of global analysis compared to the earlier ones can be surmised from
Table~\ref{tblChiSq}
\tblChiSq
which  lists the $\chi^2$'s of some of these fits, along with other comparison
parton distribution sets, for combined fixed-target and collider \smcap{DIS} as
well as \smcap{DY} data sets.\footnote{The direct photon and jet data sets are
not
included in the $\chi ^2$ table since, without including the sizable
theoretical uncertainties for the former and experimental systematic errors for
the latter, such $\chi ^2$ values do not carry the usual statistical meaning.}
We can see that almost all the change is caused by the improved accuracy of the
\smcap{HERA}.

A comparison of the inclusive jet data of \smcap{CDF} and \smcap{D0} and
results of the \smcap{CTEQ4M} parton distribution set is given in
Fig.~\ref{figJetFit}a. We see that although there is a discernible rise of the
\smcap{CDF} data above the fit curve (horizontal axis) in the high $E_t$
region, the fit agrees with the \smcap{D0} data within statistical errors
which, in turn, agree with the \smcap{CDF} data within errors. We now turn
specifically to this region since it has been the subject of much recent
attention and speculation.
\figJetFit

\section{High $E_t$ Jets and Parton Distributions}\label{sec:HiJet}

Although inclusive jet data was included in the above-described global fits, it
is understandable why the new parton distributions still under-estimates the
\smcap{CDF} cross-section: these data points have large errors,
so do not carry much statistical weight in the fitting process, and
the simple (unsigned) total $\chi^2$ is not sensitive to
the pattern that the points are uniformly higher in the large $E_t$ region.
In order to address the important question of whether the \smcap{CDF} high
$E_t$ jet data require the presence of ``new physics'', we investigated the
feasibility of accommodating these data in the conventional \smcap{QCD}
framework by exploiting the
flexibility of $G(x,Q)$ at higher values of $x$ where there are few
independent constraints, while maintaining good agreement with other data
sets in the global analysis.\cite{jet1}

To do this, it is necessary to (i) provide
enough flexibility in the parametrization of $G(x,Q_0)$ to allow for
behaviors different from the usual (but arbitrary) choice; and (ii) focus on
the high $E_t$ data points and assign them more statistical weight than
their nominal values in order to force a better agreement between theory and
experiment. Thus, the spirit of the investigation is not to obtain a ``best
fit'' in the usual sense. Rather, it is (i) to find out whether such
solutions exist; and (ii) if they do exist, to quantify how well these
solutions agree with other data sets as compared to conventional parton
distribution sets. The global analysis work described in Sec.~\ref{sec:GA}
without special attention to the high $E_t$ points provides the natural
setting to put the results of Ref.~\cite{jet1} in context.

The original study~\cite{jet1} was performed using the \smcap{CDF} Run-IA data.
 Fig.~\ref{figJetFit}b compares predictions of one of the examples obtained,
the normalization=1.0 \smcap{PDF} set (refer to here as \smcap{CTEQ4HJ}),
with the more recent Run-IB results of both \smcap{CDF} and \smcap{D0}.
For this comparison, an overall normalization factor of 1.01(0.98) for the
\smcap{CDF(D0)} data set is
found to be optimal in bringing agreement between theory and experiment.
Results shown in Table~\ref{tblChiSq} quantify the changes in $\chi^2$ values
due to the requirement of fitting the high $E_t$ jets. Compared to the best fit
\smcap{CTEQ4M}, the overall $\chi ^2$ for \smcap{CTEQ4HJ} is
indeed slightly higher.  But this difference is much smaller than the
difference between \smcap{MRSJ} and \smcap{CTEQ4M}, and those
between sets in the \smcap{CTEQ4A} series (not shown).  Thus the
price for accommodating the high $E_t$ jets is negligible.
Even though \smcap{CTEQ4HJ} does not give the absolute overall
best fit to all data, it provides an extremely good description of
all data sets. It can certainly be considered as a candidate for the gluon
distribution
if the other ones are.\footnote{%
This is to be contrasted with the conclusion of {\em incompatibility}
between the inclusive jet and \smcap{DIS} data reached by Ref.~\cite{GMRS}.
Their
fit to inclusive jet data over the full $E_t$ range (the \smcap{MRSJ'} set)
gives
rise to an extremely large $\chi ^2$ (about 20,000) for the \smcap{BCDMS} data
set.}

We will need strong, independent measurements of the
large-$x$ gluons in order to clarify the situation with the
high-$E_t$ jets. This poses theoretical as well as experimental challenges. As
an example, in order to make effective use of existing and future direct photon
production data to constrain the gluon in this region, we must understand
quantitatively the observed pattern of deviation of the $p_t$ spectrum from
\smcap{NLO QCD}, perhaps by resummation of soft gluon effects.\cite{CtqDph}

\section{Global Analysis and $\alpha_s$}

The \smcap{QCD} coupling $\alpha_s$ enters all lepton-hadron and hadron-hadron
processes. Hence, a global analysis of these processes places important
constraints on the value of $\alpha_s$ and provides a comprehensive test of its
universality. Because of the intimate coupling between $\alpha_s$ and $G(x,Q)$,
the determination of $\alpha_s$ in global analysis is not as ``clean'' as in
dedicated measurements such as \smcap{QCD} corrections to inclusive
cross-sections in $e^+e^-$ collisions or \smcap{DIS} sum rules. It, however,
provides an important complementary way to check the consistency of the theory
in a variety of different processes.

We have therefore examined the sensitivity of the various processes
participating in the global analysis to the value of $\alpha_s$, measured by
the individual $\chi^2$ of the relevant experiment or process, as we vary
$\alpha_s$ over the range (0.110, 0.122). Some results are briefly summarized
here. Fig.~\ref{figChisqAlfa}a shows the $\chi^2$'s of the \smcap{DIS}
experiments and the sum of these. Clearly seen is the known preference of the
fixed-target experiments for lower values of $\alpha_s$ in the 0.112 - 0.113
region, and the $\chi^2$'s rise rapidly for increasing $\alpha_s$. On the other
hand, the \smcap{HERA} collider data, now equally precise and numerous, show a
minimum at a higher value of $\alpha_s$ of around 0.116-0.118. These results
confirm those of the individual measurements, but in the global analysis
context where the parton distributions are constrained by a much wider set of
data involving many independent processes.  The difference in the two preferred
$\alpha_s$ ranges clearly needs to be better understood. It is possible, that
when correlated systematic errors of the experiments are fully taken into
account, this difference will become insignificant.\cite{HeraAnal} Otherwise,
one must question whether there are relevant physics which has been left out.
For instance, one may note that the collider data are concentrated in the
small-$x$ region; while the sensitivity to $\alpha_s$ of the fixed-target data
are mainly in the large-$x$ region.

\figChisqAlfa
Fig.~\ref{figChisqAlfa}b shows the nominal ``$\chi^2$'' as a function of
$\alpha_s$ for the hadronic inclusive jet data sets. The quotation marks are
used here because only statistical errors are included in the calculation at
this stage, it does not carry the statistical meaning of a true $\chi^2$. This
plot is presented here only for illustrative purposes because inclusive jets
have been advocated in the literature as a good process to measure $\alpha_s$.
We note first, the apparent difference in the minima shown by the two
\smcap{CDF} and \smcap{D0} curves using \smcap{CTEQ4A} \smcap{PDF}'s cannot be
taken seriously. This is merely a reflection of the slight differing slopes of
the respective data points seen in Fig.~\ref{figJetFit} -- some $E_t$-dependent
systematic error(s) can easily change this result. Secondly, by comparing the
two \smcap{CDF} curves obtained using \smcap{CTEQ4A} and B-series \smcap{PDF}'s
respectively, one sees the minimum of the second curve is shifted from and
steeper than that of the first.  The reason for this lies in the fact that the
constraints imposed by the jet data due to the correlation between $\alpha_s$
and $G(x,Q)$ are fully taken into account in \smcap{CTEQ4A}, but are not
included in the B-series \smcap{PDF}'s. This implies, attempted ``determination
of $\alpha_s$'' with jet data using pre-determined (``off the shelf'')
\smcap{PDF}'s will very likely give unreliable values and will certainly
under-estimate the errors.

\section{Summary}

Perturbative \smcap{QCD} phenomenology are very diverse; and the scale probed
by current experiments spans a very wide range. The overall success of
p\smcap{QCD} in providing a consistent framework for describing the physics of
all the processes included in current global analysis is quite remarkable. We
report on significant progress made in extracting the universal parton
distribution functions from these analyses due to the impact of recent
precision \smcap{DIS} and jet production data.  Of the remaining uncertainties
on $G(x,Q)$, those due to the value of $\alpha_s$ is the most significant;
hence are studies in detail.  The much discussed high $E_t$ inclusive jet
cross-section has been shown to be compatible with all existing data within the
framework of conventional p\smcap{QCD} provided flexibility is given to the
non-perturbative gluon distribution shape in the large-$x$ region. Theoretical
and experimental challenges to further clarify this issue are briefly
mentioned.

More precise knowledge of the parton distributions, as described here, not only
provides better input for the calculation of all standard model and beyond
standard model processes.  It provides a firmer basis for more definitive
probes of the region of applicability of conventional perturbative \smcap{QCD}
and for sharpening any potential contradictions. In this way, it opens the
opportunity to explore, quantitatively, the new frontiers of \smcap{QCD} as
well as ``new physics'' beyond the SM. The former are exemplified by a variety
of two- or three-large scale problems which require new resummation methods
much discussed in the theory community. Only by developing a credible
phenomenology of these new regions of phase space, can the rich content of the
quantum field theory of \smcap{QCD} -- a theory of the largest to the smallest
scales -- be fully exposed.


\begin{thebibliography}{99}
\bibitem{one}
Space limitation permits only the most directly relevant new papers be cited.
See Ref.~\cite{cteq4} for more detailed references.
\bibitem{HeraAnal}
This issue can be clarified by careful fits to all DIS data, incorporating
fully correlated errors, such as being undertaken now by the experimental
groups at HERA.
\end{thebibliography}
\end{document}